\documentclass[12pt]{article}
\usepackage{times}
\usepackage{braket}
\usepackage{amsmath}
\usepackage{slashed}
\usepackage{verbatim}
\usepackage[titletoc]{appendix}
\usepackage{graphicx}
\usepackage{cite}
\usepackage{subcaption}
\usepackage{sidecap}
\usepackage{amsthm}
\usepackage{bbm}
\usepackage{color}
\usepackage{amsthm}
\usepackage{amssymb}
\usepackage{hyperref}
\usepackage{fullpage}
\usepackage{authblk}

\begin {document}

\title{Black hole microstates vs the additivity conjectures}
\author[1]{Patrick Hayden}
\author[2]{Geoff Penington,}
\affil[1]{\small \em Stanford Institute for Theoretical Physics, Stanford University, Stanford CA 94305 USA}
\affil[2]{\small \em Center for Theoretical Physics,\\ University of California, Berkeley, CA 94720 USA}
\maketitle

\newtheorem{poss}{Possibility}
\newcommand{\Tr}{\mathrm{Tr}}

\newcommand{\cD}{\mathcal{D}}

\begin{abstract}

We argue that one of the following statements must be true: (a) extensive violations of quantum information theory's additivity conjectures exist or (b) there exists a set of `disentangled' black hole microstates that can account for the entire Bekenstein-Hawking entropy, up to at most a subleading $O(1)$ correction. Possibility (a) would be a significant result in quantum communication theory, demonstrating that entanglement can enhance the ability to transmit information much more than has currently been established. Option (b) would provide new insight into the microphysics of black holes. In particular, the disentangled microstates would have to have nontrivial structure at or outside the black hole horizon, assuming the validity of the quantum extremal surface prescription for calculating entanglement entropy in AdS/CFT.

\noindent
\footnotetext{\hspace{-0.75cm}{\tt phayden@stanford.edu, geoffp@berkeley.edu}}
\end{abstract}

\tableofcontents

\section{Introduction}
Over the last few years, ideas from quantum information have provided many new insights into quantum gravity, and in particular AdS/CFT \cite{Page:1993df, Ryu:2006bv, Hayden:2007cs, VanRaamsdonk:2010pw, Almheiri:2012rt, Almheiri:2014lwa,Brown:2015bva, Harlow:2016vwg, Penington:2019npb, Almheiri:2019psf}. Although less prevalent, there have also been examples of results from AdS/CFT leading to novel observations about quantum information \cite{Pastawski:2015qua, May:2019yxi, May:2019odp}. In this paper, we will argue instead that \emph{either} a quantum information result \emph{or} a different result about AdS/CFT must be true. Assuming that neither is true leads inevitably to a contradiction. Depending on their background, readers may be more interested in one result than the other, in which case the existence of a second possibility may be unsatisfying. However, we consider both to be potentially significant. We leave the question of deciding which one is true (or if both are) for future investigation.

\begin{poss} \label{poss:1}
Extensive violations of quantum information theory's additivity conjectures exist. 
\end{poss}
One of the earliest questions in quantum information theory, going back to the early years of information theory itself~\cite{P73}, was to calculate the rate at which classical bits can be communicated through a quantum channel $\mathcal{N}$, in the limit of a large number of channel uses. This is known as the classical capacity $C(\mathcal{N})$ of the channel. Almost forty years after the question was first posed, Holevo, Schumacher and Westmoreland showed that~\cite{holevo1998capacity,schumacher1997sending}
\begin{align} \label{eq:classcap}
C(\mathcal{N}) = \lim_{n \to \infty} \chi (\mathcal{N}^{\otimes n})
\end{align}
where the Holevo information $\chi(\mathcal{N})$~\cite{holevo1973bounds} is given by`
\begin{align} \label{eq:holevo}
\chi(\mathcal{N}) = \max_{\rho_{XA}} I(X : B)_{\mathcal{N}(\rho)}.
\end{align}
Here the channel $\mathcal{N}$ maps states on Hilbert space $A$ to states on Hilbert space $B$, while $X$ is an auxiliary \emph{classical} system. $I(X:B)$ is the quantum mutual information.

However, while correct, formula \eqref{eq:classcap} has the major defect that it involves maximising over all possible inputs to an unbounded number of copies of the channel. It therefore cannot be reliably computed in practice, even approximately. In contrast, when computing the capacity of a classical channel, it isn't necessary to take the limit $n \to \infty$. Instead, it suffices to consider a single copy of the channel.

Could this also be true for quantum channels? It is clear from the definition \eqref{eq:holevo} of $\chi$ that
\begin{align}
\chi(\mathcal{N}^{\otimes n}) \geq n \chi(\mathcal{N}).
\end{align}
The additivity conjectures proposed that this was in fact always an equality, and so $C(\mathcal{N}) = \chi(\mathcal{N})$~\cite{Holevo07}. Because equality holds provided the inputs to $\mathcal{N}^{\otimes n}$ are not entangled across channel uses, the conjectures amount to the assertion that the use of entangled codewords cannot improve the rate at which information can be transmitted through a quantum channel. 
However, it was shown by Hastings that the conjectures are false; the Holevo information can be `superadditive'~\cite{hastings2009superadditivity}. Nonetheless the known violations of additivity are small, improving capacity by less than a bit no matter how large the channel $\mathcal{N}$ is. This has continued to be true despite significant efforts over a number of years to find larger violations~\cite{brandao2010hastings,fukuda2010comments,grudka2010constructive,aubrun2011hastings,
collins2012towards,collins2013low,fukuda2014revisiting,collins2015estimates,belinschi2016almost,fukuda2017minimum}. There are some indications that larger violations could be found, however. In fact, last year, Collins and Youn found arbitrarily large violations of additivity not for tensor products of quantum channels but in the so-called commuting operator framework~\cite{collins2019superadditivity}. In finite-dimensional settings, this is equivalent to the tensor product framework, but the recent resolution of the Connes embedding and Tsirelson problems demonstrates that entanglement can behave quite differently in the two frameworks in general~\cite{ji2020mip}.

Possibility~\ref{poss:1} may appear to be remote from the alternative, Possibility~\ref{poss:2}, but we will argue that at least one of them needs to be true.

\begin{poss} \label{poss:2}
The Bekenstein-Hawking entropy of large black holes in AdS/CFT can be almost entirely accounted for by a set of `disentangled microstates'.
\end{poss}

In this case, there would have to exist a set of orthogonal black hole microstates in AdS/CFT that can account for the \emph{entire} Bekenstein-Hawking $S_{BH} $ of the black hole (up to at most an $O(1)$ subleading correction) and that are very atypical in the following sense. For a typical black hole microstate $\ket{\text{typ}}$, the entanglement entropy of a region $B$ consisting of slightly less than half the boundary will be approximately equal to the entropy of the same region in the thermal ensemble 
\begin{align}
S(\rho_\text{typ}^B) \approx S(\rho_\text{th}^B).
\end{align}
In contrast, in any of these atypical, or `disentangled', microstates, which we label $\ket{\text{disent}}$, the entanglement entropy must be much smaller. Specifically, it can be at most half the entanglement entropy of two copies of the region in the thermofield double state (TFD)
\begin{align}
S(\rho_\text{disent}^B) \lesssim \frac{1}{2} S(\rho_{TFD}^{B_L B_R}).
\end{align}
Here $B_L$ and $B_R$ represent the copies of $B$ in the left and right boundaries of the state
$\ket{TFD} \in \mathcal{H}_\text{CFT} \otimes \bar{\mathcal{H}}_\text{CFT}$ which is defined as
\begin{align} \label{eq:TFD}
\ket{TFD} \propto \sum_n e^{- \beta E_n / 2} \ket{n} \ket{\bar n}.
\end{align}
The states $\ket{n} \in \mathcal{H}_\text{CFT}$ form the energy eigenbasis, with energies $E_n$, and the states $\ket{\bar n}$ are their CPT conjugates. Each copy of the CFT contains a copy of the subregion $B$, which we label $B_L$ and $B_R$ respectively. It is, at this point, very well established that the dual bulk description of the TFD state is a two-sided AdS-Schwarzschild black hole \cite{Maldacena:2001kr}.

Since the TFD state has $O(1/G_N)$ mutual information between the two regions, the entanglement entropy $S(\rho_\text{disent}^B)$ is $O(1/G_N)$ smaller than the entanglement entropy for typical microstates, a leading order entanglement deficit. This justifies our description of these conjectural microstates as `disentangled'. To reiterate the claim we made above, in the absence of extensive violations of the additivity conjectures, there need to exist $\exp(S_{BH} - O(1))$ orthogonal disentangled microstates for all large AdS-Schwarzschild black holes.\footnote{Here, by large, we mean sufficiently large such that the mutual information between the two regions is indeed $O(1/G_N)$. The minimal horizon radius at which this is true is dimension-dependent (and in higher dimensions can only be calculated numerically), but is always comparable to the AdS-scale.} We shall often call such a set of states an `almost-basis' of disentangled microstates, because, while they do not necessarily form a basis for the full space of black hole microstates, they do necessarily form an orthonormal basis for a subspace of microstates that has approximately the same entropy as the full space. 

We emphasize, however, that a subleading $O(1)$ correction to the entropy corresponds to an $O(1)$ multiplicative change in the Hilbert space dimension.
It is completely consistent for an almost-basis of disentangled microstates to exist, and for the overlap between a typical black hole microstate and \emph{any} superposition of disentangled microstates to be small, just not for it to be parametrically small in $G_N$.

In the CFT language of entanglement entropies, the existence of disentangled microstates, even an almost-basis of disentangled microstates, may not seem terribly surprising. Suppose, rather than only considering states with some given energy (i.e. black hole microstates), we considered all states in some tensor product Hilbert space. Typical states in this space would have close to maximal entanglement entropy. However, it is easy to construct a basis for the Hilbert space that is entirely made up of product states and, hence, where every state in the basis has zero entanglement entropy. Since entanglement entropy is not a linear observable, it is not paradoxical for it to be atypically small for every single state in a complete orthonormal basis. In our case, we only need an `almost-basis', which is easier to satisfy. The ease with which a disentangled almost-basis can be constructed while disregarding the energy constraint is misleading, however. Any such product state would grossly violate the energy constraint.

Moreover, in AdS/CFT, the boundary entanglement entropy of any subregion is related via the Ryu-Takayanagi formula \cite{Ryu:2006bv, Hubeny:2007xt} and, more generally, the quantum extremal surface (QES) prescription \cite{Engelhardt:2014gca}, to the area of the minimal area bulk surface homologous to said subregion. Most famously, in the special case where boundary subregion is the entirety of one of the two boundaries in the thermofield double state, the RT formula tells us that the entanglement entropy is equal to the Bekenstein-Hawking entropy $S_{BH} = A_\text{hor}/4 G_N$, where $A_\text{hor}$ is the area of the black hole horizon.

\begin{figure}[t]
\includegraphics[width = 0.35\linewidth]{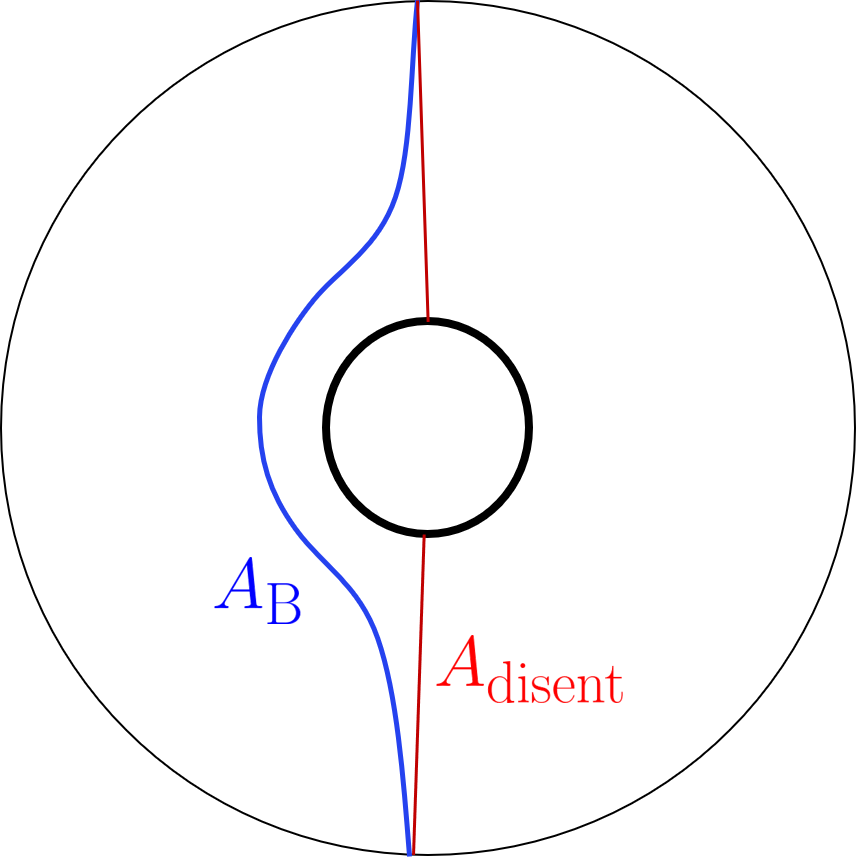}
\centering
\caption{Using the Ryu-Takayanagi formula, one finds that the typical black hole microstate entropy is given by the area $A_B$ of the blue surface (the minimal area surface homologous to $B$ in the black hole static slice). The disentangled microstate entropy is upper bounded by the area $A_\text{disent}$ of the red surface (the minimal area surface homologous to the union of $B$ and part of the bifurcation surface).}
\label{fig:difference}
\end{figure}

For the subregion $B$ that we are interested in, the surface whose area determines the entanglement entropy for typical microstates, as well as the surface determining the maximal entropy for disentangled microstates, are shown in Figure \ref{fig:difference}. Unsurprisingly, the area of the typical RT surface is simply too large to give the entanglement entropy needed for a disentangled microstate.

Moreover, at least naively, the area of the Ryu-Takayanagi surface should be a linear observable in the bulk theory. (See, for example, the discussions in \cite{Almheiri:2016blp, Harlow:2016vwg}.) If true, this would suggest that there exists an entire \emph{typical subspace} containing almost all microstates within which the RT surface area always ensures a typical entanglement entropy. The existence of such a typical subspace would completely rule out the possibility of an almost-basis of disentangled microstates.

Of course, it is easy to construct states with the same energy as an AdS-Schwarzschild black hole where the area of the RT surface (and hence the entanglement entropy) is significantly smaller. Simple examples are given by configurations of ordinary matter that will eventually collapse to black hole. However, the entropy of states of ordinary matter in AdS cannot come close to saturating $S_{BH}$, scaling instead as $S_{BH}^{d/(d+1)}$ in $(d+1)$-dimensional bulk spacetimes~\cite{hooft1993dimensional}. They cannot account for an almost-basis of disentangled microstates.

Instead, there appear to be three scenarios that would allow Possibility~\ref{poss:2} to hold. The first is that the disentangled microstates have no bulk interpretation, and can only be described in the language of the boundary CFT. This would mean that AdS/CFT is a not true duality between two different theories and is instead merely an alternative gravitational description for some \emph{subset} (or maybe code subspace) of CFT states. The second is that the disentangled microstates have a bulk description, but that the QES prescription does not apply to this bulk description. In this case it seems important to understand why the usual derivation of the QES prescription using the replica trick \cite{Lewkowycz:2013nqa, Faulkner:2013ana, Dong:2017xht, Almheiri:2019qdq, Penington:2019kki} fails in these disentangled states. Finally, the disentangled microstates could have a bulk description in which the QES prescription applies, but where the area of the Ryu-Takayanagi surface (or minimal QES) is much smaller than in typical states. The issue here, as we saw in the discussion of ordinary matter configurations above, is not finding any bulk states with a much smaller RT surface, but finding enough orthogonal states to produce an almost-basis.

In particular, if this last possibility is true, we will argue that the geometries associated to the disentangled microstates would have to have some sort of structure (i.e. some difference from an ordinary AdS-Schwarzschild black hole) at or outside the event horizon of the black hole. Possible examples of such structure are fuzzball geometries~\cite{Mathur:2005zp, Mathur:2008nj, Raju:2018xue}, a firewall~\cite{Almheiri:2013hfa, Almheiri:2012rt} or an end-of-the-world brane~\cite{VanRaamsdonk:2010pw, Hartman:2013qma, Cooper:2018cmb}. However, it is an open question whether they can potentially give the almost-basis of disentangled states that we need.

Explaining the entire Bekenstein-Hawking entropy of a black hole using states that have structure at the horizon is the explicit goal of the fuzzball programme, and is closely related to the firewall conjecture. However, advocates of both these points of view tend to believe that all states, or at least all typical states, have structure at the horizon. We emphasize that this is far more than is required to avoid extensive additivity violations. Instead it is sufficient for structure to exist merely for a particular, very atypical almost-basis of states. 

If there is a linear observable that measures whether there is structure at the horizon, this would mean that typical states would also have to have an $O(1)$ probability of having structure at the horizon. Alternatively, however, we could imagine that there exist nonperturbatively small overlaps between states with different geometries, allowing states with smooth horizons to be written as a superposition of exponentially many states with structure at the horizon. Such a view was previously advocated prominently in \cite{Marolf:2020xie} for example. It implies that an operator measuring whether there is structure at the horizon must be state dependent \cite{Papadodimas:2012aq}, or more precisely code-subspace dependent \cite{Almheiri:2014lwa, Hayden:2018khn}.

\subsection*{Connecting the possibilities}

The link between Possibilities 1 and 2 is to define a quantum channel $\mathcal{N}$ mapping a code subspace of AdS/CFT black hole microstates to a portion $B$ of the CFT boundary. A na\"ive analysis of that channel leads to extensive violations of the additivity conjectures. Careful scrutiny of the argument, however, identifies a gap, which corresponds to Possibility 2.

The structure of the paper is as follows. In Section \ref{sec:additivity}, we review the additivity conjectures. In Section \ref{sec:paradox}, we explain the contradiction between the absence of extensive violations of additivity and the absence of an almost-basis of disentangled microstates. In Section \ref{sec:structure}, we argue that, assuming that they have a geometrical interpretation and obey the QES prescription, the disentangled microstates must have structure at, or outside, the horizon. Finally we conclude in Section \ref{sec:discuss} with a review of the arguments from the paper and a few general comments.

\section{The additivity conjectures} \label{sec:additivity}

The additivity of the Holevo information $\chi(\mathcal{N})$ was the first of several additivity conjectures in quantum information that were ultimately demonstrated to be equivalent~\cite{Pomeransky03,AudenaertB04,MatsumotoSW04,shor2004equivalence}. The simplest of these is the additivity of the minimum output entropy of a quantum channel~\cite{KingR01}, which is the form we will investigate. Let $\cD(A)$ be the space of density operators on a Hilbert space $A$ and $S(\rho)$ the von Neumann entropy of the density operator $\rho$. The minimum output entropy of a quantum channel $\mathcal{N} : \cD(A) \rightarrow \cD(B)$ is defined to be
\begin{equation}
S_{\min}(\mathcal{N}) = \min_{\rho \in \cD(A)} S(\mathcal{N}(\rho)).
\end{equation}
By the concavity of the entropy, the minimum is always achieved by a pure state. The additivity conjecture states that for two channels $\mathcal{N}_j : \cD(A_j) \rightarrow \cD(B_j)$,
\begin{equation} \label{eq:additivity}
S_{\min}(\mathcal{N}_1 \otimes \mathcal{N}_2)  = S_{\min}(\mathcal{N}_1)  + S_{\min}(\mathcal{N}_2).
\end{equation}
When the states $\rho_j$ are minimizers for the corresponding $\mathcal{N}_j$, the state $\rho_1 \otimes \rho_2$ is a valid input to $\mathcal{N}_1 \otimes \mathcal{N}_2$, ensuring that the lefthand side of \eqref{eq:additivity} is always bounded above by the right. The nontrivial direction of the conjecture was that the opposite inequality also held, that is, that the output entropy of a product channel could always be minimized by a product state. The product state $\rho_1 \otimes \rho_2$ is in fact a local minimum of the output entropy~\cite{gour2012minimum}, but Hastings found examples for which it is not a \emph{global minimum}~\cite{hastings2009superadditivity}. Since then the largest gap ever demonstrated is an underwhelming
\begin{equation}
\Delta := S_{\min}(\mathcal{N}_1)  + S_{\min}(\mathcal{N}_2) - S_{\min}(\mathcal{N}_1 \otimes \mathcal{N}_2)
\end{equation}
of $\log 2$~\cite{belinschi2016almost}.\footnote{Technically, a sequence of channels has been identified with gaps converging to $\log 2$ from below.}
Thus, the additivity conjecture is false, but in a thoroughly unsatisfying way. Exact evaluation of the capacity of a quantum channel requires the consideration of entangled codewords and the evaluation of an intractable limit, but the capacity improvements are of little practical importance -- at most a bit for all known examples, regardless of the size of the channels in question.\footnote{As mentioned earlier, however, larger violations have been found in a relaxed setting where tensor products of quantum channels are replaced with a commutation condition~\cite{collins2019superadditivity}. 
In a celebrated advance, the interchangeability of tensor products and commutation conditions for studying bipartite quantum correlations was shown to be false earlier this year~\cite{ji2020mip}, a question known as Tsirelson's problem which is equivalent~\cite{junge2011connes} to the Connes embedding conjecture in operator algebras~\cite{connes1976classification}. 
If we are primarily interested in the increase in the classical capacity of a channel from the use of entangled codewords, it is crucial to consider genuine tensor products of channels, rather than merely commuting subalgebras. More generally, working within the commuting subalgebra framework without imposing significant restrictions can potentially yield unphysical results~\cite{summers1990independence}.}

The limited nature of the known additivity violations is in sharp contrast to the analogous question for R\'enyi entropies
\begin{equation}
S_\alpha(\rho) = \frac{1}{1-\alpha} \log \Tr \rho^\alpha
\quad \text{with} \quad
S_{\alpha, \min}(\mathcal{N}) = \min_{\rho \in \cD(A)} S_\alpha(\mathcal{N}(\rho)).
\end{equation}
In that case, for all $\alpha > 1$ there are examples in which the gap is roughly $\log \dim A_1 = \log \dim A_2$ for arbitrarily large $A_j$~\cite{H07,hayden2008counterexamples}. For these examples, the minimum output R\'enyi entropy of the product channel $\mathcal{N}_1 \otimes \mathcal{N}_2$ is almost the same as the minimum output entropy of either $\mathcal{N}_1$ or $\mathcal{N}_2$ alone. 
Gaps of that size persisting in the von Neumann entropy limit $\alpha \rightarrow 1$ would correspond to the existence of channels that are almost useless for communication using product codewords but can send bits at half the rate of a \emph{noiseless} channel when entangled codewords are permitted. Precisely that behavior is even known to occur for the \emph{private} capacity of a quantum channel, which characterizes the rate at which secret key bits can be established between the sender and receiver~\cite{smith2009extensive}.

No known theorems exclude the possibility that the use of entangled codewords could have an extensive impact on the classical capacity of a channel. Indeed, that is exactly our
Possibility~\ref{poss:1}, which can be stated more precisely as follows. There exist infinite sequences of channels $$\mathcal{N}_j^{(n)} : \mathcal{S}(A_j^{(n)}) \to \mathcal{S}(B_j^{(n)}),$$ with $n \subseteq \mathbb{N}$ and $j\in\{1,2\}$, such that the Hilbert spaces $A_j^{(n)}$ and  $B_j^{(n)}$ grow linearly with $n$ in qubit terms, and for which the gap $\Delta^{(n)}$ satisfies
\begin{equation}
\Delta^{(n)} \geq Cn
\end{equation}
for some constant $C > 0$.

\section{A contradiction} \label{sec:paradox}
In this section, we show that the absence of extensive violations of the additivity conjectures implies the existence of an almost-basis of disentangled microstates. 

In the semiclassical limit where Newton's constant $G_N$ is small (or equivalently the gauge group rank $N$ is large), the thermofield double state is concentrated in an energy window of width $O(\sqrt{1/G_N})$ about an $O(1/G_N)$ saddle point energy (in AdS units). We can therefore approximate the thermofield double state with high accuracy, even if we restrict the sum over energy eigenstates in \eqref{eq:TFD} to this energy band. This gives an approximate thermofield double state $\ket{TFD'} \in \mathcal{H}_\text{code} \otimes \bar{\mathcal{H}}_\text{code}$, where $\mathcal{H}_\text{code} \in \mathcal{H}_\text{CFT}$ is the subspace spanned by energy eigenstates $\ket{n}$ in the energy window. Explicitly,
\begin{align}
\ket{TFD'} \propto \sum_{\ket{n} \in \mathcal{H}_\text{code}} e^{- \beta E_n / 2} \ket{n} \ket{\bar n}.
\end{align}

We now consider the channel $\mathcal{N}: \cD(\mathcal{H}_\text{code}) \to \cD(\mathcal{H}_B)$ that maps states in the code space $\mathcal{H}_\text{code}$ to their restriction to the boundary region $B$, which we shall take to be slightly less than half of the boundary. For this to be a well-defined finite-dimensional quantum channel, we would need to regularise the CFT on some lattice. So long as we take the regularisation scale to be much smaller than the black hole inverse temperature, this seems unlikely to cause any problems.

For a generic state $\rho \in \cD(\mathcal{H}_\text{code})$, the von Neumann entropy $S(\mathcal{N}(\rho))$ is given by the Ryu-Takayanagi formula as  $A_B/4G_N$ (plus subleading corrections) where $A_B$ is the area of the minimal area bulk surface homologous to the boundary region $B$.
In contrast, the von Neumann entropy
\begin{align}\label{eq:violation}
S\left(\mathcal{N} \otimes \bar{\mathcal{N}}(\ket{TFD} \bra{TFD})\right) 
= S(\rho_{TFD}^{B_L B_R}) = \frac{A_\text{disent}}{2 G_N} < \frac{2 A_B}{4 G_N}.
\end{align}
because the minimal area surface homologous to $B_L B_R$ passes through the Einstein-Rosen bridge rather than being disconnected. Here, we are defining $A_\text{disent}$ to be half the area of the minimal surface homologous to $B_L B_R$, or equivalently, to be equal to the area of the minimal surface homologous to $B_R$ plus some subregion of the bifurcation surface, as in Figure \ref{fig:difference}.

To see how a contradiction will emerge, let us temporarily assume that for \emph{every} state $\rho \in \cD(\mathcal{H}_\text{code})$, the output entropy satisfies $S(\mathcal{N}(\rho)) = A_{B}/4G_N$. In other words, we assume that every state in the code space is a typical black hole microstate. This assumption is obviously untrue as stated; we shall refine the argument and weaken the required assumption below.

Given our assumptions, the minimum output entropy of a single copy of the channel $\mathcal{N}$ would be $A_{B}/4G_N$. However, with two copies of the channel, we can input the approximate the thermofield double state, with output entropy only $A_\text{disent}/2G_N$. Crucially the difference between $2 \times A_{B}/4G_N$ and $A_\text{disent}/2G_N$ grows extensively in the limit of small $G_N$.

Moreover, the von Neumann entropy satisfies Fannes' inequality, which states that for any two quantum states $\rho, \sigma \in \cD(\mathcal{H})$, we have
\begin{align}
\left| S(\rho) - S(\sigma) \right| \leq  \lVert \rho - \sigma \rVert_1 \log \dim \mathcal{H} + \eta( \lVert \rho - \sigma \rVert_1 ),
\end{align}
where $\eta(\epsilon)$ vanishes with $\epsilon$.
A finite-dimensional regularisation of the CFT Hilbert space $\mathcal{H}_\text{CFT}$ will have $\log \dim \mathcal{H}_\text{CFT} = O(1/G_N)$. Hence, so long as the trace distance between our approximate TFD state $\ket{TFD'}$ and the exact thermofield double state $\ket{TFD}$ tends to zero in the semiclassical limit (which it does by construction), the von Neumann entropy of the two states will agree at leading order. This implies an extensive violation of the additivity of minimum output entropy.

Of course, the assumption that every state in the code space $\mathcal{H}_\text{code}$ (as currently constructed) has output entropy $A_{B}/4G_N$ is clearly untrue. The energy window will, for example, contain states whose bulk dual does not contain a black hole at all, and instead simply contains a low density cloud of radiation spread over a very large region.

However, such states are expected to be `rare'; `typical' states in the code space will indeed have entropy $A_{B}/4G_N$. Since we only want to approximate the TFD state, one might hope that we could just `throw away' bad states until we end up with a code space where the minimum output entropy is close to $A_{B}/4G_N$. 

More concretely, we can simply repeatedly find the state in the code space with minimum output entropy, and then restrict our code space to the orthogonal complement of this minimum entropy state. If we are to avoid an extensive violation of the additivity of minimum output entropy, this procedure cannot increase the minimum output entropy above $A_\text{disent}/4G$ (at leading order), until we have removed so many states that the TFD state can no longer be approximated by two copies of the code space with an error that tends to zero (no matter how slowly) in the semiclassical limit $G_N \to 0$.
 
In other words, there needs to exist a projective measurement which, when applied to the thermofield double state (or equivalently the one-sided thermal ensemble), will output a `disentangled microstate' with $O(1)$ probability. Since the probability of projecting onto any single state (with energy equal to the saddle point energy) in the thermal ensemble is $e^{-S_{BH}}$, it seems that we would have to have $O(e^{S_{BH}})$ distentangled microstates within the code subspace. Equivalently, the thermodynamic entropy of the set of disentangled black hole microstates would have to be equal to the Bekenstein-Hawking entropy of the black hole minus, at most, an $O(1)$ subleading correction. Otherwise there must be an extensive violation of the additivity conjectures.

Unfortunately, the argument above is a bit too quick -- to rigorously reach the same result will require a bit more work. The problem is that the probabilty of a given energy eigenstate appearing in the thermal ensemble is only $O(e^{-S_{BH}})$ if the energy of the eigenstate is equal to the saddle point energy, plus at most an $O(1)$ correction. However, the energy window that we are considering needed to have width $O(\sqrt{N})$ in order to be able to approximate the TFD. If our disentangled microstates all happened to have lower energy than the saddle point energy, then we could have avoid extensive violations with only $e^{S_{BH} - O(\sqrt{1/G_N}})$ disentangled microstates.

It is intuitively obvious why this does not avoid the existence of an almost-basis of disentangled microstates: if the disentangled microstates all have lower energy than the saddle point energy, they may not form an almost-basis for our original black hole, but they will form an almost-basis for a slightly lower temperature black hole. Suppose we have found an orthonormal set of disentangled microstates, with associated subspace projector $P_\text{disent}$, such that
\begin{align} \label{eq:pdisent}
p^{(\beta_0)}_\text{disent} = \frac{\Tr(P_\text{disent} e^{-\beta_0 H})}{\Tr(e^{-\beta_0 H})} = O(1),
\end{align}
for some black hole temperature $\beta_0$. This is exactly the condition that we found above was needed to avoid extensive additivity violations. We can then vary the temperature of the black hole in order to maximise $p^{(\beta)}_\text{disent}$. From now on, we will use $\beta$ to refer specifically to this maximising temperature. Setting the derivative of \eqref{eq:pdisent} to zero and rearranging terms, we have
\begin{align} \label{eq:energyopt}
\frac{\Tr(P_\text{disent} H e^{-\beta H})}{\Tr(P_\text{disent} e^{-\beta H})} = \frac{\Tr(H e^{-\beta H})}{\Tr( e^{-\beta H})}.
\end{align}

We will now show that the number of disentangled microstates $\text{Rank}(P_\text{disent})$ is equal to the Bekenstein-Hawking entropy $S_{BH}$ of a black hole at temperature $\beta$, minus at most an $O(1)$ correction. In other words, we will make rigorous the heuristic argument given above. We have
\begin{align}
\text{Rank}(P_\text{disent}) = \Tr(P_\text{disent}) = \sum_{\ket{n}} \braket{n|P_\text{disent}|n} = \sum_n P_{nn},
\end{align}
where $\ket{n}$ refers to the energy eigenbasis as before. If we define the normalised probability distribution $q_n = P_{nn} e^{-\beta E_n} / \Tr(P_\text{disent} e^{-\beta H})$ then
\begin{align}
\frac{\text{Rank}(P_\text{disent})}{\Tr(P_\text{disent} e^{-\beta H})} =  \sum_n q_n e^{\beta E_n} \geq \Tr(P_\text{disent} e^{-\beta H}) e^{\beta \sum_n q_n E_n} = e^{\beta \Tr(P_\text{disent} H e^{-\beta H})/ \Tr(P_\text{disent} e^{-\beta H})}.
\end{align}
Here the inequality follows by Jensen's inequality for the convex function $e^x$. Finally, we use \eqref{eq:energyopt} to remove the projectors $P_\text{disent}$ from the exponent and get
\begin{align} \label{eq:finalcount}
\text{Rank}(P_\text{disent}) \geq \Tr(P_\text{disent}\, e^{-\beta H}) \,e^{\beta \Tr( H e^{-\beta H})/ \Tr( e^{-\beta H})} = p^{(\beta)}_\text{disent}\, e^{S_{BH}}.
\end{align}
In the final equality, we used the Gibbs state entropy formula $S_{BH} = \beta \braket{H} + \log Z$. Since $p^{(\beta)}_\text{disent} = O(1)$ by assumption, \eqref{eq:finalcount} is exactly our desired result.

\section{The QES prescription and structure at the horizon} \label{sec:structure}

We showed above that the absence of extensive violations of the additivity violations implies the existence of a set of `disentangled microstates' that can account for almost the entire Bekenstein-Hawking entropy. In this section, we argue that, if these disentangled microstates a) have a geometrical bulk description, and b) have entanglement entropies that obey the QES prescription, then they must have structure at or outside the horizon. We then briefly discuss possible sources of such structure include firewalls, fuzzballs and end-of-the-world branes, as well as the analogue of disentangled microstates in tensor network toy models.

\subsection{Disentangled microstates must have horizon structure}
The QES prescription\cite{Ryu:2006bv, Hubeny:2007xt, Faulkner:2013ana, Engelhardt:2014gca} says that the entropy $S(B)$ is given by\footnote{One should also include the effects of higher-curvature corrections \cite{Dong:2013qoa}.}
\begin{align}
  S(B)  = \min \text{ext}_\gamma \left[ \frac{A(\gamma)}{4 G} + S_\text{bulk}(\gamma) \right].
\end{align}
That is, we first look for all surfaces $\gamma$ homologous to $B$ that are extrema of the \emph{generalised entropy} -- the area $A(\gamma)$ of the surface $\gamma$ in Planckian units plus the entropy $S_\text{bulk}(\gamma)$ of the matter fields between $\gamma$ and $B$. These are known as quantum extremal surfaces (QES). The entropy $S(B)$ is then given by the generalised entropy of the minimal QES (the QES with smallest generalised entropy).

For our purposes it will be more convenient to work with an alternative formulation of the same rule, known as the maximin prescription \cite{Wall:2012uf, Akers:2019lzs}
\begin{align}
  S(B)  = \max_C \min_{\gamma \in C} \left[ \frac{A(\gamma)}{4 G} + S_\text{bulk}(\gamma) \right].
\end{align}
In this version of the rule, one first looks for the minimal generalised entropy surface $\gamma$ within some particular AdS-Cauchy slice $C$, and the maximises this minimum over all possible choice of Cauchy slice. Although often less practical for explicitly finding the minimal QES than just looking directly for extremal surfaces, an advantage of this formulation is that it allows us to put bounds on the entanglement entropy without having to actually find any QES at all.\footnote{For this reason, the maximin prescription is vital for proving important properties of the QES prescription such as nesting and strong sub-additivity.}  In particular, if we can lower bound the generalised entropy of \emph{all} surfaces within \emph{some} Cauchy slice, we automatically lower bound the generalised entropy of the minimal QES.

There are known examples of states where the QES prescription fails (even at leading order), as discussed in \cite{Akers:2020pmf}. However, in these known cases, the true boundary entropy is still lower-bounded by the area of the minimal area surface (within the maximal Cauchy slice) homologous to $B$. The large violations of the QES prescription simply come from the bulk entropy term (which competes with the area term at leading order in the states in question) giving a smaller contribution than the na\"ive QES calculation would suggest. In this case, we consider only the contribution from classical area in calculating our lower bound, and so such effects should not impact our argument.

Suppose that the microstate geometry contains the entire exterior patch of the AdS-Schwarzschild spacetime, with metric
\begin{align}
 ds^2 = - f(r) dt^2 + f(r)^{-1} dr^2 + r^2 d \Omega^2.
\end{align}
The detailed form of $f(r)$ will not be important, but we note that $f(r_\text{hor}) = 0$ and $f'(r_\text{hor}) = 4 \pi/\beta $ where $\beta$ is the inverse temperature of the black hole. We will assume throughout this paper that the black hole is large, with $r_\text{hor} \gtrsim 1$ in AdS units; for small black holes $S(\rho_{TFD}^{B_L B_R}) = 2 S(\rho_{th}^B)$ at leading order and there is no need for disentangled microstates to avoid extensive violations of additivity.

We make no assumptions about what happens behind the horizon, but presumably the disentangled microstate geometry ends somewhere in the interior, either smoothly, or at some stringy object such as an end-of-the-world brane. We can then consider a Cauchy slice $C$ that looks like a static AdS-Schwarzschild slice everywhere in the exterior region. Within this Cauchy slice, the minimal area surface, homologous to the union of $B$ and any subregion of the bifurcation surface, has area \begin{align}
A_\text{disent} = 4 G_N S(\rho_{TFD}^{B_L B_R})/2
\end{align}
(up to $O(1)$ corrections from the $S_\text{bulk}$ term). This is because we can always glue together to copies of such a surface at the horizon to get a surface homologous to $B_L B_R$ within the static slice of two-sided AdS-Schwarzschild.
\begin{figure}[t]
\includegraphics[width = 0.35\linewidth]{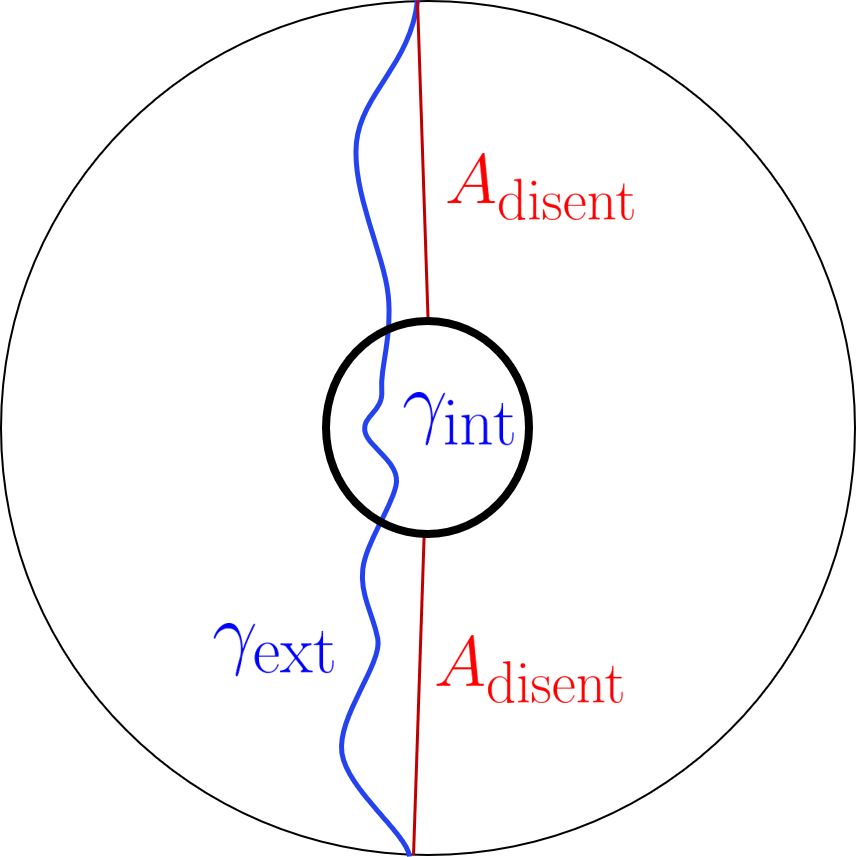}
\centering
\caption{A schematic of the Cauchy slice $C$, which looks like AdS-Schwarzschild outside the event horizon. Any surface $\gamma$ homologous to $B$ splits into an exterior part $\gamma_\text{ext}$ and an interior part $\gamma_\text{int}$ (which may be empty). The area of the exterior part $\gamma_\text{ext}$ is bounded from below by $A_\text{disent}$. Since the area of the minimal area surface in $C$ lowers bounds the generalised entropy of the minimal QES, the entanglement entropy of $B$ is lower bounded by $A_\text{disent}/4 G_N$.}
\label{fig:maximin}
\end{figure}

Now consider the intersection with the exterior patch of any surface $\gamma$ in $C$ that is homologous to $B$, as shown in Figure \ref{fig:maximin}. The resulting surface $\gamma_\text{ext}$ will be homologous to the union of $B$ and some part of the bifurcation surface, and hence has area at least $A_\text{disent}$. Since $\gamma_\text{ext} \subseteq \gamma$, the same is true of $\gamma$. Since the area term lower bounds the generalised entropy,\footnote{One might worry about issues of renormalisation here, since (depending on your choice of renormalisation scheme) the renormalised bulk entropy can be negative. However, this still won't affect the entropy at leading order unless Planckian modes are disentangled, which would create a firewall.} the generalised entropy 
\begin{align}
S_\text{gen}(\gamma) \geq \frac{1}{2}S(\rho_{TFD}^{B_L B_R}).
\end{align}
To make this inequality strict, we note that the only surface $\gamma_\text{ext}$ that saturates the area lower bound is homologous to (almost) half the bifurcation surface (in addition to $B$). So long as any surface $\gamma$ that is a) homologous only to $B$ and b) contains the minimising surface $\gamma_\text{ext}$ must have some additional area outside the exterior patch, then the inequality will be strict. This will be true so long as some smooth geometry continues beyond the bifurcation surface. (Since we previously neglecting $O(1)$ corrections to the generalised entropy, we need this additional area to be more than Planckian in size.)

Of course, the assumption that the entire exterior patch looks like AdS-Schwarzschild is a strong one. It is untrue, for example, for black holes formed from collapse, even a long time $t$ after they formed. Instead the AdS-Schwarzschild approximation breaks down, within the static slice, at $r = r_\text{hor}[1 + O( e^{-4 \pi t/\beta})]$, very slightly \emph{outside} the horizon. So can black holes formed from collapse always have small $S(B)$? The answer, of course, is no; $S(B)$ quickly grows after the black hole forms until it reaches the equilibrium thermal value.

Our lower bound on $S(B)$ simply wasn't very good because we made a bad choice of Cauchy slice. Instead, we should have used the fact that the black hole formed from collapse looks like AdS-Schwarzschild both in the exterior region \emph{and} in the black hole region, so long as we look at infalling times long after the black hole formation. 

In Eddington-Finkelstein coordinates, the AdS-Schwarzschild metric for these regions is
\begin{align}
ds^2 = - f(r) dv^2 + 2 dv dr + r^2 d\Omega^2.
\end{align}
This time, we shall assume that the spacetime looks like AdS-Schwarzschild so long as a) the infalling time $v > v_\text{min}$ and b) the radius $r > r_\text{min}$. Here $r_\text{min}$ is a small, but not parametrically small, distance inside the horizon, while $v_\text{min} \ll - O(\beta)$ is much earlier than the boundary time ($v = 0$) where we measure $S(B)$. 

We consider a Cauchy slice that looks like the static slice for $r > r_\text{out}$ (the exterior segment), for some $r_\text{out}$ a small, but again not parametrically small, distance outside the horizon. This lies at an infalling time $v_\text{out} = - O(\beta)$.  We then attach a lightlike piece that lies at constant infalling time and at radii $r_\text{min} \leq r \leq r_\text{out}$ (the lightlike segment). Finally, we attach a constant radius piece that covers infalling times $v_\text{min} \leq v_\text{out}$ at  (the interior segment). 

By similar arguments to those above, the minimal area of the intersection $\gamma_\text{ext}$ of the surface $\gamma$ with the exterior segment is $A_\text{disent}$, minus a correction that is parametrically smaller than the black hole horizon area $A_\text{hor} = O(r_\text{hor}^{d-1})$. More precisely, by linearising $f(r)$ near the horizon we find that the correction is $O\left(r_\text{hor}^{d-2}\sqrt{\beta (r_\text{out} - r_\text{hor})}\right)$. Since $\beta \lesssim r_\text{hor}$ for AdS-Schwarzschild black holes, this is parametrically smaller than the horizon area, as claimed.

Moreover, the area of $\gamma_\text{ext}$ can only be smaller than $A_\text{disent}$ if $\gamma_\text{ext}$ is homologous to roughly half the sphere at radius $r_\text{out}$ (together with $B$). Similarly, if the area of the lightlike segment $\gamma_\text{light}$ of $\gamma$ is also to be parametrically smaller than the horizon area, then the interior segment $\gamma_\text{int}$ must be homologous to roughly half the sphere at $r_\text{min}$. However, the area of the interior segment is then at least the minimum of $O(r_\text{hor}^{d-1})$ and $O(r_\text{hor}^{d-1} (v_\text{out} - v_\text{min})\sqrt{(r_\text{hor} - r_\text{min})/\beta})$, since  it either has to cover half the sphere at $(r_\text{min},v_\text{out})$ or to connect the sphere at $(r_\text{min},v_\text{out})$ to a sphere at $(r_\text{min},v_\text{min})$. This means that the minimal area of $\gamma$ is strictly bigger than $A_\text{disent}$ so long as $v_\text{out} - v_\text{min} \gg \beta$ and $r_\text{hor} - r_\text{min} \gtrsim r_\text{out} - r_\text{hor}$. Given our original assumptions about $r_\text{min}$ and $v_\text{min}$, we can always choose $r_\text{out}$ and $v_\text{out}$ so that this is true.

In summary, if the disentangled microstates are geometrical and obey the QES prescription, they cannot either a) have a spacetime geometry that contains the entire exterior patch of AdS-Scharzschild, together with any smooth geometry beyond the bifurcation surface, or b) a spacetime geometry that looks like the exterior and black hole interior of AdS-Schwarzschild for all radii greater than $r_\text{min} < r_\text{hor} - O(r_\text{hor})$ and infalling times greater than $v_\text{min} \ll -O(\beta)$. In other words, the atypical microstates must have some structure (or at least some geometry that differs from AdS-Schwarzschild) at or outside the horizon.

\subsection{Possible sources of horizon structure}
Having argued that disentangled microstates need to have structure at the horizon, we should also comment on what that structure might be. One possibility is a \emph{fuzzball}, a horizonless geometry supported by string degrees of freedom that looks like a black hole to observers who remain outside the Schwarzschild radius \cite{Mathur:2005zp, Mathur:2008nj, Raju:2018xue}. Although the end goal of the fuzzball program is to show that all black hole microstates are really fuzzballs,\footnote{One reason to be sceptical about the strongest versions of this vision is the success of the two-sided black hole in explaining properties of the thermofield double state \cite{Maldacena:2001kr}, including effects such as traversable wormholes \cite{Gao:2016bin} that had not previously been known from the boundary description. This strongly suggests that horizons can exist, at least in some low-energy `effective' description of the physics (that may involve some coarse-graining or ensemble-averaging).} it has so far only been possible to account for the \emph{entire} black hole entropy using fuzzballs for `small' black holes, where the classical horizon area is zero. Moreover, the best understood fuzzball states are either extremal or near-extremal black holes, not the Schwarzschild black holes that we care about in this paper. Such black holes have a very long throat which prevents the existence of large amounts of mutual information between subregions on each side of a two-sided black hole, and so the additivity arguments are not relevant for them. Finally, even if we could construct fuzzball AdS-Schwarzschild black holes, it is not clear whether the QES prescription would actually then give the correct answer for a disentangled microstate.

A (closely related) possibility is a \emph{firewall}, a high-energy `wall' at the horizon where the spacetime ends. Firewalls were first conjectured to appear in evaporating black holes (at or before the Page time) in \cite{Almheiri:2012rt} in order to avoid monogamy of entanglement paradoxes. More general arguments that firewalls should exist in \emph{typical} black hole microstates were given in \cite{Almheiri:2013hfa}; again the justification was to avoid paradoxes. Since then there have been a number of arguments that the firewall paradoxes can be avoided via the $A=R_B$ \cite{Nomura:2012cx} or $ER=EPR$ proposal \cite{Maldacena:2013xja}, the state-dependence of operator reconstruction \cite{Papadodimas:2012aq} and the complexity of experimentally realising a paradox \cite{Harlow:2013tf}. These ideas got significant support from the quantum extremal surface and replica wormhole calculations of \cite{Penington:2019npb, Almheiri:2019psf, Penington:2019kki,Almheiri:2019qdq}. Nonetheless significant conceptual issues remain in understanding how a smooth horizon can be consistent with unitarity. We emphasize however that there have never been compelling dynamical arguments for the emergence of a firewall. Semiclassical gravity always favours the emergence of smooth horizons. Instead, the arguments for firewalls have always been based on trying to show that they are the \emph{only} possibility consistent with unitarity.
\begin{figure}[t]
\includegraphics[width = 0.17\linewidth]{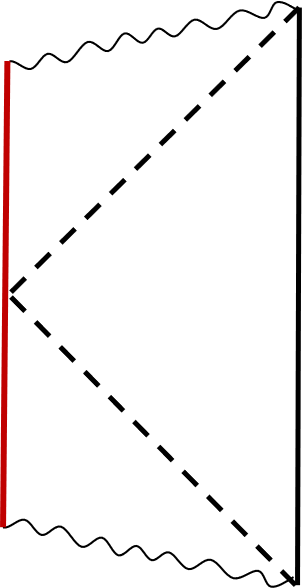}
\centering
\caption{Penrose diagram for a $\mathbf{Z}_2$ quotient of the two-sided black hole, an example of a spacetime with the correct properties to be an disentangled microstate.}
\label{fig:penrose}
\end{figure}

A more explicit construction of a state that has exactly the right entanglement entropy to be an disentangled microstate is the $\mathbf{Z}_2$ quotient of a two-sided AdS-Schwarzschild black hole \cite{Hartman:2013qma}. In this state we identify the left and right halves of the two-sided black hole, creating a state that ends on an end-of-the-world brane passing through bifurcation surface as shown in Figure \ref{fig:penrose}. The Ryu-Takayanagi surface is allowed to end on this brane, so the RT formula gives exactly the required value for an disentangled microstate. Note that unlike the fuzzballs and firewalls discussed above, such a state does have a smooth horizon (for both the black and the white hole); it just doesn't have a smooth bifurcation surface. If we time evolve the state either forwards or backwards in time, we find that the entanglement entropy grows until it saturates at the entanglement entropy for a typical microstate. This is consistent with the idea that disentangled microstates should be highly atypical, even if they do end up forming an almost-basis. Time evolution will tend to cause them to thermalise. It has been speculated that states with an end-of-the-world brane at the bifurcation surface (or more generally at any RT surface) should be able to account for the entire Bekenstein-Hawking entropy \cite{VanRaamsdonk:2010pw}. However, no explicit construction is known for the exponentially large number of states that would be necessary. One approach would be to start with an arbitrary product state in the regularised CFT, and then evolve in Euclidean time. However this seems likely to typically produce something closer to a Kourkoulou-Maldacena state, with an end-of-the-world brane far behind the black hole horizon.\footnote{We thank Douglas Stanford for discussions on this point.}
\begin{figure}[t]
\begin{subfigure}{.48\textwidth}
  \centering
 \includegraphics[width = 0.8\linewidth]{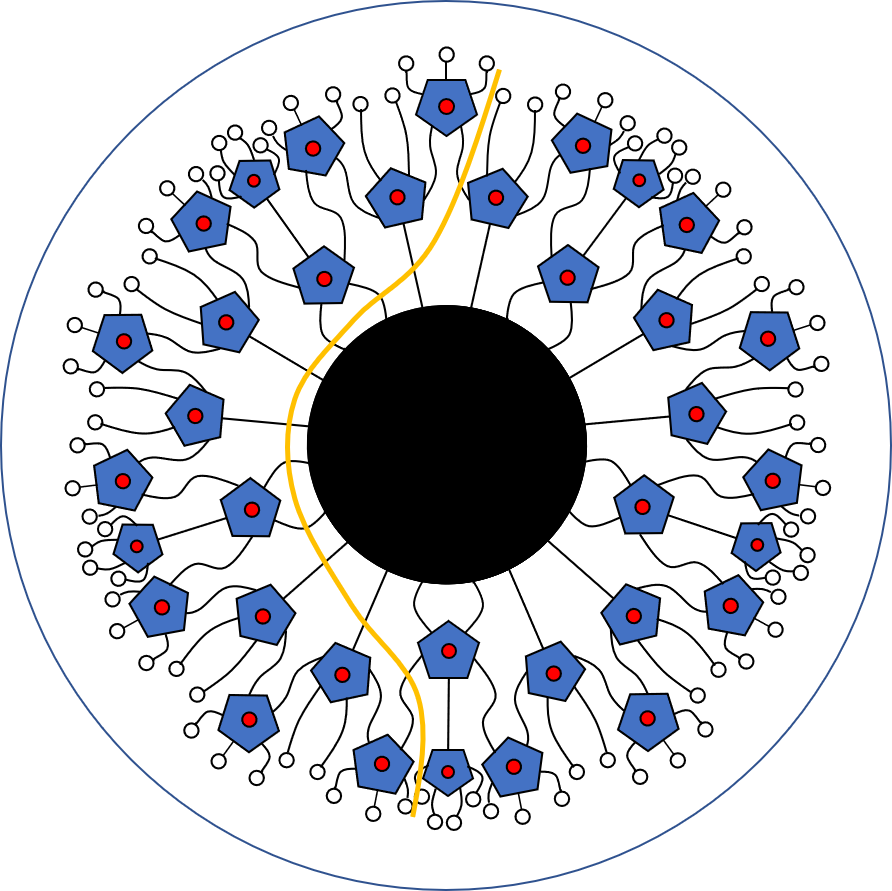}
\end{subfigure}
\begin{subfigure}{.48\textwidth}
  \centering
 \includegraphics[width = 0.8\linewidth]{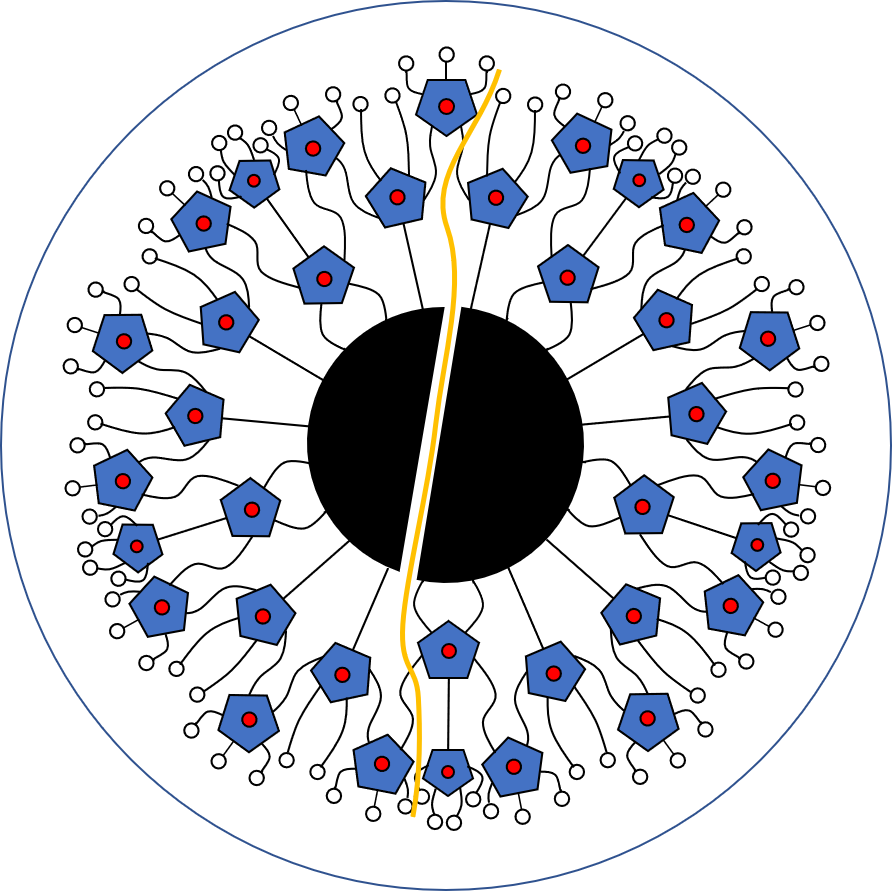}
\end{subfigure}
\centering
\caption{Tensor networks analogues of a typical microstate (left) and a disentangled microstate (right). The  disentangled microstate has a product tensor inserted at the horizon legs, in contrast to the typical microstate which has a Haar random tensor. The minimal cuts determining the entanglement entropies are shown in orange.}
\label{fig:tensor}
\end{figure}

Arguably the strongest reason to believe in the existence of a disentangled basis of black hole microstates comes not from gravity itself, but from random tensor network toy models of AdS/CFT \cite{Swingle:2009bg, Pastawski:2015qua, Hayden:2016cfa}. A typical black hole microstate is conventionally modelled in such a tensor network by a very large Haar random tensor (the black hole) surrounded by a network of smaller tensors (the exterior geometry), as shown in Figure \ref{fig:tensor}. A two-sided black hole is modelled by gluing together two copies of the exterior network at their horizons. Disentangled microstates would naturally correspond to states where a product tensor (rather than a random tensor) is inserted at the horizon. It is easy to see that such states have the correct entanglement entropy (by looking at the minimal cut through the network). They also can easily be used to create a basis of microstates, in the sense that any Haar random tensor ``typical microstate'' can be written as a superposition of these ``disentangled microstates''. However, there are a few reasons to be cautious about concluding that analogous disentangled microstates must exist in gravity. Firstly, the disentangled tensor network states will almost certainly not lie within the small energy band required for them to be actual black hole microstates in AdS/CFT. Secondly, tensor networks are only thought to be a good analogy for AdS/CFT at scales at least as large as the AdS scale. In contrast, the disentangled microstates needed to avoid extensive additivity violations would have to have structure a Planckian distance from the horizon.

In summary, there are a number of possibilities for structure at the horizon. However, there is no compelling gravitational argument that sufficiently many disentangled microstates should exist (for AdS-Schawrzschild black holes) in order to be able to account for essentially the entire Bekenstein-Hawking entropy. The fact that an almost-basis of such disentangled microstates is implied by the absence of extensive additivity violations can be viewed as new evidence in favour of their existence.

\section{Summary and Discussion} \label{sec:discuss}
In this paper, we have argued that either:
\begin{itemize}
\item \textbf{Possibility~\ref{poss:1}}: Extensive violations of the additivity conjectures exist,  or 
\item \textbf{Possibility~\ref{poss:2}}: An almost-basis of disentangled microstates must exist in AdS/CFT. 
\end{itemize}
Our argument worked by constructing an approximation for the thermofield double state by entangling two `code subspaces' of black hole microstates. We then imagined repeatedly excluding states from the code subspace if their entanglement entropy (for some particular boundary subregion $B$) was less than (or comparable to) half the TFD entanglement entropy for the two-sided subregion $B_L \cup B_R$. 

If we continued this process for long enough, eventually we would either a) run out of such states, while still being able to make a good approximation to the TFD out of two copies of the code subspace, or b) stop being able to approximate the TFD. The former would imply the existence of extensive violations of minimum output entropy (and hence of all the additivity conjectures). In the latter case, we would have a orthonormal set of  $O(e^{S_{BH}})$ states (the ones that we excluded from the code subspace) whose entanglement entropy is at most half the TFD entanglement entropy $S(B_L B_R)$. These would form the almost-basis of disentangled microstates.

Finally, we speculated on what these disentangled microstates might look like if they exist. We argued that, if a) they had a geometrical bulk interpretation (at least outside the horizon) and b) obeyed the RT formula/QES prescription, then they must have some sort of nontrivial structure at or outside the horizon. We noted that such states can naturally be constructed in tensor network toy models, and commented on the potential relationship with various proposed forms of horizon structure, including fuzzballs, firewalls and end-of-the-world branes.

Let us end by making a couple of comments on the significance and consequences of each possibility being true. If Possibility~\ref{poss:2} is true, there is an obvious new target for trying to understand quantum black holes: finding these exponentially many disentangled microstates. Moreover, such a possibility also emphasizes how hard it is to create short wormholes out of entangled black holes. A common thought experiment is to imagine Alice and Bob flying to distant galaxies with a large number of qubits, which they use to create a wormhole and meet in the middle. To avoid extensive violations of the additivity conjectures, their same method of black hole construction would have to be capable of making exponentially many different disentangled microstates (if Bob measured his qubits in the right basis) if the wormhole is ever to be short enough to have large mutual information (and hence for them to meet).

It is interesting in this context to recall from Section \ref{sec:additivity} that extensive violations of  the minimum output R\'enyi entropies for $\alpha > 1$ are easy to find. In gravity, this would correspond to a short wormhole threaded by a positive tension cosmic brane. It seems intuitive that a positive tension brane can help shorten a wormhole. However it would be nice to understand this more precisely.

In contrast, if Possibility~\ref{poss:1} is true, then constructing classical codes for quantum channels without using entangled codewords is not even approximately optimal. This leaves us with a giant elephant in the room: how to construct better codes and how (if possible) to efficiently bound the capacity that can be gained by using entanglement. There is an ongoing effort to improve upper bounds on the capacity~\cite{leditzky2018approaches,wang2019converse,fang2019geometric,ding2020bounding}. For cases in which the best existing bounds have been explicitly evaluated, they appear to approximate the capacity quite closely in practice. Trying to evaluate them on the channels described here could be illuminating.

\section{Acknowledgements}
We would like to thank Benoit Collins, Steve Shenker, Douglas Stanford, Lenny Susskind, Michael Walter and Mark Wilde for valuable discussions. This work was supported by  AFOSR award FA9550-19-1-0369, CIFAR, DOE award DE-SC0019380 and the Simons Foundation.

\small
 \bibliographystyle{unsrt}
\bibliography{biblio}
\end{document}